\documentclass{article}
\pdfoutput=1
\usepackage[dvipsnames]{xcolor}
\usepackage[utf8]{inputenc}
\usepackage{svg}
\usepackage{caption}
\usepackage{graphicx}
\usepackage[english]{babel}
\usepackage{amsmath}
\usepackage{hyperref}
\usepackage{breakurl}
\usepackage[T1]{fontenc}
\usepackage{multicol}
\usepackage{mathtools}
\usepackage{float}
\usepackage{amssymb}
\usepackage{xfrac}
\usepackage{doi}
\usepackage[margin=0.75in]{geometry}
\usepackage[maxbibnames=99,sorting=none]{biblatex}
\usepackage{amsthm}
\usepackage{ytableau}
\usepackage{braket}
\usepackage{csquotes}
\usepackage{titling}
\AtBeginBibliography{}
\newcommand{\be}{\begin{equation}}
\newcommand{\ee}{\end{equation}}

\usepackage{caption}
\usepackage{graphicx}
\usepackage{subfiles}
\usepackage{abstract}
\usepackage[makeroom]{cancel}
\usepackage{tensor}
\usepackage{youngtab}

\numberwithin{equation}{section}

\begin{document}

\date{\vspace{-5ex}}
\title{%
{\textbf{Computing Nonequilibrium Transport from Short-Time Transients: From Lorentz Gas to Heat Conduction in One Dimensional Chains}}\\[2ex]
\normalsize Davide Carbone\\
\footnotesize
\textit{Laboratoire de Physique de l’Ecole Normale Supérieure, ENS Université PSL, CNRS, Sorbonne Université, Université de Paris, Paris, France}\\
\footnotesize davide.carbone@phys.ens.fr\\[1.5ex]
\normalsize Vincenzo Di Florio\\
\footnotesize
\textit{MOX Laboratory, Department of Mathematics, Politecnico di Milano, Piazza Leonardo Da Vinci, 32, 20133 Milano, Italy}\\
\footnotesize
\textit{and CONCEPT Lab, Fondazione Istituto Italiano di Tecnologia, Via E. Melen 83, Genova, 16152, Italy}\\
\footnotesize vincenzo.diflorio@polimi.it\\[1.5ex]
\normalsize Stefano Lepri\\
\footnotesize
\textit{Consiglio Nazionale delle Ricerche, Istituto dei Sistemi Complessi, Via Madonna del Piano, 10, 50019, Sesto Fiorentino, Italy}\\
\footnotesize
\textit{and INFN, Sezione di Firenze, Via G. Sansone 1, 50019, Sesto Fiorentino, Italy}\\
\footnotesize stefano.lepri@cnr.it\\[1.5ex]
\normalsize Lamberto Rondoni\\
\footnotesize
\textit{INFN, Sezione di Torino, Via P. Giuria 1, 10125 Torino, Italy}\\
\footnotesize
\textit{and Dipartimento di Scienze Matematiche, Politecnico di Torino, Corso Duca degli Abruzzi 24, 10129 Torino, Italy}\\
\footnotesize lamberto.rondoni@polito.it
}
\leftskip=0cm
\rightskip=0cm
\maketitle
\vspace{-8ex}
{\footnotesize\noindent\textbf{Corresponding authors}: Davide Carbone, davide.carbone@phys.ens.fr, and
Vincenzo Di Florio, vincenzo.diflorio@polimi.it }\\
\noindent\rule{\textwidth}{1pt}
\section*{\large Abstract}
\small We test the Transient Time Correlation Function (TTCF) method to 
compute nonequilibrium transport coefficients, highlighting its conceptual and practical difference from standard time–average approach. While time averages extract transport properties from long stationary trajectories and discard transient dynamics, TTCF adopts the complementary strategy: it exploits the information contained in short-time transients following the onset of an external perturbation, while discarding the long-time evolution once stationarity is reached.\\
We revisit the theoretical framework of TTCF and assess its numerical performance through representative case studies, the Lorentz gas and
a many-body system, namely a 
chain of oscillators with anharmonic pinning
potential. By direct comparison with time averages, we show that for the Lorentz Gas TTCF yields consistent transport coefficients in both linear and nonlinear regimes at a reduced computational cost. Moreover, the TTCF displays superior precision in the linear-response regime, and remains reliable in non-ergodic situations, revealing the presence of regions of phase space corresponding to different behaviors, as well as the possibility of phase transitions. 
For the anharmonic chain,  we show that the TTCF is a scalable and efficient alternative for the numerical study of nonequilibrium transport.
\newline
\newline
\textit{Key words}: nonequilibrium transport, response theory, transient time correlation function, ergodicity breaking
\newline

\section{\large Introduction}


%
One task of nonequilibrium statistical mechanics is to make predictions about the evolution of small or strongly driven systems, whose statistical properties and collective behaviour are hard to compute. The first case comes under the topic of linear response theory, investigated by Onsager, Casimir and Kubo, cfr. \cite{onsager1931reciprocal,onsager1931reciprocal2,Casimir,kubo1966fluctuation}, in presence of time reversal symmetry, or less strict hypotheses \cite{robnik1986false, carbone2022time}. On the other hand, the exact theory of response to perturbations of any magnitude, has been developed just recently within the field of nonequilibrium molecular dynamics \cite{evans2008fluctuation}; it is also known as TTCF formalism because based on a {\it transient time correlation function}, and has proven useful in numerous applications in which phase-space statistics is hard to obtain \cite{maffioli,MAFFIOLI2024109205,bernardi2016local,talaei2012local}. At the same time, other methods have been proven more efficient in treating Langevin type processes \cite{spacek2025transient}. Although Langevin processes are substantially different from deterministic ones in many situations, the range of applicability and efficiency of TTCF is an important question to investigate.

When considering steady state properties of the systems of interest, one may choose different approaches. The most obvious, from a numerical point of view, is the time average of an observable along a phase space trajectory, also because in statistical mechanics that is postulated to represent the result of a macroscopic measurement \cite{mejia2025nonequilibrium}.
Then, provided the relevant invariant probability distribution exists and the system is ergodic with respect to it,  time averages coincide with the (presumably) much handier ensemble averages, and the result of a measurement becomes a property of this distribution. A perturbation, however, changes this picture, as the state, hence the probability distribution  supposedly representing it, start changing in time, and the connection between the two becomes dubious, unless a number of conditions are met, including a vast separation of microscopic and time time scales \cite{DiFlorio}. 
\textcolor{black}{This time dependence is problematic for small systems,
{\color{black} for systems subject to very strong drivings, or evolving too fast,}
but the equivalence between the macroscopic (or collective) state and the associated probability distribution is expected to be restored once a new steady state is reached.} The difficulty may be that the final distribution is not known, or, as typical of deterministic dissipative systems, is singular and hard to handle analytically.
\textcolor{black}{Therefore, one could choose to perform direct time averages, which, however, do not by themselves provide a systematic theoretical description. Moreover, this strategy entails several difficulties: it requires very long simulations to control fluctuations and, in the presence of slow relaxation or multiple dynamically relevant regions, it may yield ambiguous results.}
{\color{black} Also, at times one would like to characterize the transients states, and not just the stationary states.}

Linear response theory provides a reliable alternative, if a system is evolving close to equilibrium, avoiding situations such as phase transitions. In this respect, the TTCF formalism constitutes a tool of much wider applicability, although the physical interpretation of its results, being purely probabilistic, may require some care.
Given the initial ensemble, typically representing an equilibrium state, TTCF produces the average of an observable at time $t$ after the perturbation, using the time integral of the transient correlation between the observable and the so-called dissipation function 
$\Omega^{(0)}$. This is determined by the unperturbed distribution $f_0$, and by the phase-space volumes variation rate of the dynamics. It encodes the dissipation associated with the perturbation. Under the correlations decay 
condition that allows, for instance, the Fluctuation Relations to hold \cite{searles2007steady}, 
convergence to a stationary distribution is guaranteed, hence the probabilistic asymptotic nonequilibrium response is guaranteed to exist
\cite{typic}. 
The TTCF bypasses the explicit construction of the invariant measure of the perturbed dynamics and will be shown here to reveal possible ergodicity breakings.
This becomes transparent in the Lorentz-gas considered here, where certain driving fields create regions in the phase space corresponding, respectively, to positive and to vanishing currents.

One issue is the efficiency of the TTCF formalism results in producing the response.
Our aim is, on the other hand, methodological: we compute the response of selected observables using TTCF and compare it with estimates obtained from direct time averaging of the perturbed dynamics, precisely in those regimes where convergence is slow, or the phase space is effectively fragmented. A particularly telling fact emerges in the weak-forcing regime, in which the TTCF can be dramatically more precise than linear response or direct  time-averaging. This precision is not obtained for free, but it can be obtained at competitive computational cost, since the method trades extremely long single-trajectory averages for moderately long ensemble evolutions in the unperturbed setting.

The rest of the paper is organized as follows. In Sec.~2 we recall the dynamical-systems framework---flows, time-reversal invariance, observables, and the role of ergodicity---and introduce the TTCF response identity and the dissipation function. We then specify the two models and the observables of interest, emphasizing the features (thermostats, explicit $f_0$, possible coexistence of attracting structures) that make them suitable testing grounds for response theory. In Sec.~3 we present the numerical evidence comparing TTCF predictions with time-averaged measurements, highlighting when and why the correlation-based approach is advantageous.
\textcolor{black}{\subsection*{Related Works}
The transient time correlation function (TTCF) has a long history in nonequilibrium statistical mechanics. Early applications focused on isothermal systems and linear transport, establishing the formalism for connecting microscopic dynamics with macroscopic fluxes \cite{morriss1985isothermal,morriss1987application,evans1988transient,j2008statistical}. Subsequent work extended the approach to more complex phenomena, including synchronization transitions in coupled systems \cite{amadori2022exact} and response to time-dependent perturbations \cite{petravic1997nonlinear,iannella2023exact}. More recently, TTCF has been applied in quantum contexts \cite{greppi2025quantum} and studied from the perspective of computational efficiency, exploring accelerated algorithms and optimized implementations for large-scale simulations \cite{todd1997application,hartkamp2012transient,maffioli,MAFFIOLI2024109205}. More recently, TTCF and related response approaches have been used to study confined systems and finite-size effects, highlighting the role of system size on transport properties and the computation of viscosities in molecular systems \cite{bernardi2012response,bernardi2015system}. Collectively, these works demonstrate both the versatility of TTCF as a tool for nonequilibrium analysis and the ongoing interest in improving its practical applicability.
}

\section{\large Theory}
Consider a dynamical system described by the ordinary differential equation
\begin{equation}
\dot{\Gamma} = G(\Gamma), \qquad \Gamma \in \mathcal{M}\;,
\end{equation}
where $\mathcal{M} \subseteq \mathbb{R}^n$ denotes the phase space and $\Gamma$ represents the microscopic state of the system.

Let
\begin{equation}
\begin{split}
\Phi^t : \mathcal{M} &\rightarrow \mathcal{M}, \\
\Gamma &\mapsto \Phi^t(\Gamma)\;,
\end{split}
\end{equation}
be the time-evolution operator, such that $\Phi^t(\Gamma)$ is the phase-space point reached at time $t$ starting from the initial condition $\Gamma$. The family of maps ${\Phi^t}_{t \in \mathbb{R}}$ defines a flow on $\mathcal{M}$, which uniquely determines the trajectory of the system in phase space and fully characterizes its dynamics.

In the following, we restrict our attention to systems that are invariant under time reversal (TRI), such as Hamiltonian systems, Gaussian thermostatted dynamics, Nosé–Hoover dynamics, and related models. Time-reversal invariance implies the existence of an involutive transformation $\mathcal{I} : \mathcal{M} \to \mathcal{M}$ satisfying
\begin{equation}
\mathcal{I} \circ \Phi^t = \Phi^{-t} \circ \mathcal{I}\;,
\end{equation}
which ensures that every trajectory has a time-reversed counterpart consistent with the microscopic equations of motion.

Depending on the properties of the vector field $G(\Gamma)$, the system may exhibit regular, chaotic, or mixed behavior. Here, we focus on 
{\color{red} mainly}
chaotic dynamical systems, for which a statistical description {\color{black} is necessary} despite the deterministic nature of their dynamics.

Macroscopic properties of the system are described in terms of observables, defined as 
{\color{black} integrable}
functions on phase space,
\begin{equation}
O : \mathcal{M} \rightarrow \mathbb{R}\;,
\end{equation}
which assign a real value to each microscopic state. The macroscopic value of an observable corresponding to an initial condition $\Gamma \in \mathcal{M}$ is defined through its infinite-time average,
\begin{equation}
\overline{O}(\Gamma) = \lim_{T \to \infty} \overline{O}_T(\Gamma)\;,
\label{eq:time_av}
\end{equation}
with
\begin{equation}
\overline{O}_T(\Gamma) = \frac{1}{T} \int_0^T O\bigl(\Phi^t(\Gamma)\bigr)\, \mathrm{d}t.
\end{equation}
Typical observables include energies, particle currents, and collective coordinates.
If the limit in Eq.~\eqref{eq:time_av} exists, 
{\color{black}
it can be obtained from a suitable invariant probability measure $\mu$.
If, in addition,} its value is independent of the initial condition,
{\color{black} except for those in a set of vanishing $\mu$-measure,}
the system is said to be ergodic with respect to $\mu$. 
In this case, {\color{black} one can write:}
\begin{equation}
\overline{O}(\Gamma) = \int_{\mathcal{M}} O \, \mathrm{d}\mu \equiv \mathbb{E}_\mu[O]\;,
\end{equation}
for $\mu$-almost every $\Gamma$, meaning that possible violations are confined to a set $E \subset \mathcal{M}$ of vanishing $\mu$-measure. Under these conditions, time averages coincide with ensemble averages, establishing a fundamental link between microscopic dynamics and macroscopic observables.

Having established this equivalence, we now address the response of macroscopic observables to external perturbations. In nonequilibrium statistical mechanics, a central problem is to understand how the statistical properties of a dynamical system are modified when its equations of motion are perturbed.\\
In general, for the perturbed dynamics neither the invariant probability measure nor the ergodic properties of the system are known a priori. Assuming that the system relaxes toward a nonequilibrium steady state, one may compute the observable value by performing the time average in Eq.~\eqref{eq:time_av} along a single trajectory of the perturbed system. However, {\color{black} as mentioned above, this approach suffers from several limitations. In particular, 
it may require exceedingly long 
simulations, or it may yield incorrect results, due to the 
lack of ergodicity.}
To overcome these difficulties, response theories based on correlation functions provide a powerful alternative. Near equilibrium, fluctuation–dissipation relations connect the linear response of a system to external perturbations with suitable correlation functions computed in the unperturbed steady state, characterized by a probability density $f_0$
\cite{kubo1966fluctuation,j2008statistical,tuckerman2023statistical,mejia2025nonequilibrium}. As these relations are restricted to weak perturbations, they naturally motivate more general response formulations.
Within linear response theory, the variation of a macroscopic observable can be expressed in terms of {\color{black} the equilibrium correlation function of the evolved flows and the driving, where the evolution is the one dictated by the unperturbed (equlibrium) dynamics. This result applicability is restricted to small perturbations. }

{\color{black}
In the case of the TTCF formalism, one also computes a correlation function with respect to the unperturbed probability distribution $f_0$, but the correlated functions are the observable evolved with respect to the {\it perturbed} dynamics and the so-called dissipation function $\Omega^{(0)}$.}
{\color{black} Interestingly}, the TTCF formalism yields an exact expression for the response, valid arbitrarily far from equilibrium, provided the initial ensemble is known, {\color{violet} which is usually the case}~\cite{evans2008fluctuation,typic}. {\color{black} As in this framework the response of the system is entirely determined by correlations computed with respect to the unperturbed distribution, one does not need to}
sample the unknown steady state of the perturbed dynamics.
Specifically, the phase-space average of an observable $O$ at time $t$ after the perturbation can be written as
\begin{equation}
    \mathbb{E}_t[O] = \mathbb{E}_0[O] +\int_0^t \mathbb{E}_0\left[\Omega^{(0)}(O\circ\Phi^s)\right] \, \mathrm{d}s \;,
    \label{eq:ttcf}
\end{equation}
where $\mathbb{E}_0[\cdot]$ denotes the phase-space average with respect to the initial (equilibrium) distribution, while $\mathbb{E}_t[\cdot]$ refers to the ensemble evolved up to time $t$ under the perturbed dynamics. The integrand on the right hand side is called {\it transient time correlation function}, precisely because it has an analogue expression to a two point correlation function at equilibrium, being however computed along an out-of-equilibrium dynamical evolution $\Phi^t$
{\color{black} that does not preserve $f_0$.} The dissipation function
\begin{equation}
    \Omega^{(0)}(\Gamma) = -G(\Gamma)\nabla_{\Gamma} \ln{f_0}\Big|_{\Gamma} - \nabla_{\Gamma}\cdot G\Big|_{\Gamma}
    \label{eq:diss_func}
\end{equation}
and the initial phase-space measure 
is defined by
$\mathrm{d}\mu_0(\Gamma) = f_0(\Gamma)\,\mathrm{d}\Gamma$.

The TTCF formula is among the few exact response relations available for systems arbitrarily far from equilibrium. It is particularly relevant for small or strongly driven systems, where the assumptions of local thermodynamic equilibrium do not apply. Equation~\eqref{eq:ttcf} describes the response of an ensemble of identically prepared systems initially distributed according to $f_0$ and subsequently evolving under the perturbed dynamics generated by $\Phi^t$.\\
If the ensemble averages $\mathbb{E}_0\left[\Omega^{(0)}(O\circ\Phi^s)\right]$ decay faster than $O(1/s)$, the long-time limit can be taken, yielding
\begin{equation}
    \mathbb{E}_\infty[O] = \mathbb{E}_0[O] +\int_0^\infty \mathbb{E}_0\left[\Omega^{(0)}(O\circ\Phi^s)\right] \, \mathrm{d}s \;,
    \label{eq:ttcf_inf}
\end{equation}
which provides the asymptotic response of the system to the applied perturbation. \textcolor{black}{ The existence of the integral in Eq.\eqref{eq:ttcf_inf}, and consequently of the left-hand side, defines the condition known as $\Omega$t-mixing, which, unlike standard mixing, applies to transient states and can express relaxation to stationary states in terms of loss of time correlations with macroscopic states. It also constitutes a necessary and sufficient condition for relaxation in the sense of ensembles, while mixing is only sufficient; cf.\ Refs. \cite{mejia2025nonequilibrium,evans2016fundamentals,evans2016typicality} for more details. Being necessary and sufficient, it represents the relevant ergodic notion in our work.
}

\section{Models}

In the following, we apply the TTCF formalism to two paradigmatic models: the Lorentz gas and a 
chain of oscillators with 
anharmonic pinning (or substrate) potential. 
For both system, we compute the response of selected observables using the TTCF approach and systematically compare the results with those obtained from direct time averaging of the perturbed dynamics. This comparison allows us to assess the effectiveness of the TTCF framework, particularly in regimes where long relaxation times, multimodality, or slow convergence hinder the reliability of time-averaged measurements.

\subsection{Lorentz gas}
The Lorentz gas was first introduced in 1905 by Lorentz~\cite{lorentz1905motion} and has since been extensively studied due to its conceptual simplicity and its ability to capture essential features of transport and nonequilibrium statistical mechanics~\cite{lebowitz1978transport,lloyd1994breakdown,lloyd1995nonequilibrium,dettmann2000lorentz}. 
In its simplest formulation, the dynamics describes a particle of unit mass undergoing free motion between elastic collisions with fixed convex scatterers. 
Despite its minimal ingredients, the Lorentz gas exhibits strong chaotic behavior and provides a paradigmatic example of deterministic diffusion.

The model considered in this work is a modified Lorentz gas~\cite{lloyd1994breakdown,lloyd1995nonequilibrium}, consisting of an infinite two-dimensional triangular lattice of hard scatterers through which a single point particle moves. The wandering particle undergoes elastic hard-core collisions with the scatterers and is subject to the combined action of an external field and a thermostat.
Collisions with the scatterers are instantaneous and elastic, resulting in specular reflection of the velocity. To prevent unbounded energy growth and to ensure the existence of a nonequilibrium steady state, the dynamics is supplemented with a Gaussian isokinetic thermostat, which enforces constant kinetic energy. The equations of motion in the presence of an external field applied in the negative $x$-direction, $\mathbf{E} = (-E_x,0)$, read
\begin{equation}
\dot{\mathbf{r}} = \mathbf{v}, \qquad
\dot{\mathbf{v}} = \mathbf{F} + \mathbf{E} - \alpha(\Gamma)\mathbf{v}\;,
\label{eq:lorentz}
\end{equation}
where $\mathbf{r} = (x,y)$ and $\mathbf{v} = (v_x,v_y)$, while $\mathbf{F} = (F_x,F_y)$ denotes the impulsive force due to collisions with the scatterers. The thermostat multiplier $\alpha(\Gamma)$ is determined by the isokinetic constraint $\mathrm{d}(\mathbf{v}^2)/\mathrm{d}t = 0$, yielding
\begin{equation}
\alpha(\Gamma) = \frac{\mathbf{F}\cdot\mathbf{v} + \mathbf{E}\cdot\mathbf{v}}{\mathbf{v}^2}.
\end{equation}
This dynamics is deterministic, time-reversal invariant, and dissipative in phase space, 
{\color{black} in the sense that it does not preserve phase space volumes, and on average it contracts them. Consequently, the invariant probability distribution is singular, supported on a set of vanishing volume.}
A natural observable of interest in this model is the particle current along the direction of the applied field, defined as
\begin{equation}
\label{eq:flux_lorentz}
J(\Gamma) = \mathbf{v}\cdot \hat{\mathbf{E}},
\end{equation}
where $\hat{\mathbf{E}}$ is the unit vector parallel to the external field. The macroscopic response of the system is characterized by the change in the average current induced by the field.

It has been shown that, for certain values of the applied field, the phase space 
undergoes bifurcations leading to the emergence of {\color{black}
separate invariant sets} with non-vanishing measure~\cite{lloyd1994breakdown}. At the level of real-space dynamics, this corresponds,
{\color{black} in particular, to the existence of}
trajectories confined to bounded regions, which may sustain either vanishing or non-vanishing particle currents. In such regimes, time averages computed along individual trajectories may yield zero current even in the presence of a non-zero external field, depending on the specific attractor reached.
This phenomenon highlights a fundamental limitation of time averaging in multimodal dynamical systems. As will be shown in the results section, this issue is naturally overcome within the TTCF framework. By expressing the response as an ensemble average over a multitude of trajectories, weighted according to their phase-space measure, the TTCF correctly accounts for the coexistence of dynamically distinct regions and yields a non-zero macroscopic current whenever such contributions are statistically relevant.

For the isokinetic Lorentz gas considered here, the dissipation function entering the TTCF formalism can be computed explicitly. Since the unperturbed equilibrium measure is uniform on the constant kinetic energy surface, the associated phase-space density $f_0$ is constant along the energy shell. As a consequence, the gradient term in Eq.~\eqref{eq:diss_func} vanishes,
\begin{equation}
\nabla_\Gamma \ln f_0(\Gamma) = 0\;,
\end{equation}
and the dissipation function reduces to minus the phase-space divergence of the flow,
\begin{equation}
\Omega^{(0)}(\Gamma) = - \nabla_\Gamma \cdot G(\Gamma)\;.
\end{equation}
For the equations of motion Eq.~\eqref{eq:lorentz},
the phase-space divergence receives contributions only from the velocity components. The impulsive collision force $\mathbf{F}$ and the external field $\mathbf{E}$ do not contribute to the divergence, while the thermostat term yields
\begin{equation}
\nabla_\Gamma \cdot G(\Gamma) = - 2\alpha(\Gamma) -\nabla_\Gamma\alpha(\Gamma) \,\cdot \mathbf{v} = -\frac{E_xv_x}{\mathbf{v}^2}\;.
\end{equation}
Therefore, for the Lorentz gas with Gaussian isokinetic thermostat, the dissipation function is directly proportional to the instantaneous power injected by the external field, normalized by the kinetic energy. This quantity measures the local phase-space contraction rate and provides the microscopic source term governing the system’s response within the TTCF formalism.

\subsection{Anharmonic chain}
\label{sec:pinned}
As a prototypical example of an extended many-body system, we consider a one-dimensional  
oscillator chain with a nonlinear pinning potential~\cite{lepri2003thermal,lepri2016heat}. The system consists of $N$ particles of mass $m$, coupled by linear nearest-neighbor interactions and subject to an on-site pinning potential. 
The total potential energy of the system is given by
\begin{equation}
V = \sum_{i=1}^{N-1} \frac{1}{2}\left(x_{i+1} - x_i - a\right)^2+	\sum_{i=1}^{N} \left[ \frac{1}{2}(x_i - i a)^2 + \frac{1}{4}(x_i - i a)^4 \right]\;,
\label{eq:pot_pin}
\end{equation}
where $x_i$ denotes the position of the $i$-th particle and $a$ is the equilibrium lattice spacing. The first term represents harmonic nearest-neighbor interactions, while the remaining terms correspond to a quartic on-site pinning potential.
It is convenient to introduce the displacement from equilibrium, $r_i = x_i - i a$, which fully characterizes the dynamics, since the absolute positions $x_i$ are redundant once the equilibrium lattice is fixed. In terms of $r_i$, the force associated with the pinning potential reads

\begin{equation}
F_{\mathrm{pin}}(r_i) = -\frac{\mathrm{d}}{\mathrm{d} r_i}
\left( \frac{1}{2} r_i^2 + \frac{1}{4} r_i^4 \right)
= -\left( r_i + r_i^3 \right).
\end{equation}
Accordingly, the equations of motion for the bulk particles ($i = 2, \dots, N-1$) can be written directly in terms of the displacements as
\begin{equation}
\dot{r}_i = \frac{p_i}{m}, \qquad
\dot{p}_i = (r_{i+1} + r_{i-1} - 2r_i) + F_{\mathrm{pin}}(r_i)\;.
\end{equation}


It is known that the presence of pinning breaks translational invariance and suppresses momentum conservation, significantly affecting transport and relaxation properties. Heat transport is diffusive 
with a size-dependent thermal conductivity
$\kappa(N)$ converging to a finite value
in the large $N$ limit \cite{hu2000heat,aoki2000bulk}.

To drive the system out of equilibrium, Nosé–Hoover thermostats are applied to the first and last mobile particles. The equations of motion for the boundary particles become
\begin{align}
    \dot{p}_1 &= (r_{2} + r_{0} - 2r_1) + F_{\mathrm{pin}}(r_1) - \Psi_L p_1, \\
    \dot{p}_N &= (r_{N+1} + r_{N-1} - 2r_N) + F_{\mathrm{pin}}(r_N) - \Psi_R p_N,
\end{align}
where $\Psi_L$ and $\Psi_R$ are the Nosé–Hoover thermostat variables associated with the left and right boundaries, respectively, that evolve according to
\begin{equation}
\dot{\Psi}_{L,R} = \frac{1}{\theta^2_{L,R}}\left( \frac{p^2_{1,N}/m}{T_{L,R}} - 1 \right),
\end{equation}
where  $T_{L,R}$ are the imposed temperatures, and $\theta_{L,R}$ are the thermostat relaxation times. The dynamics defines a deterministic, time-reversal invariant flow on $\mathcal{M}$. Due to the presence of thermostats, the dynamics is dissipative in phase space and admits nonequilibrium steady states characterized by non-zero energy fluxes.

In this situation, the invariant phase-space probability density associated with the Nosé–Hoover dynamics is known
explicitly~\cite{todd2017nonequilibrium} and is given by
\begin{equation}
f_0(\mathbf{x},\mathbf{p},\Psi_L,\Psi_R)
\propto
\exp\left[
-\beta
\left(
H(\mathbf{x},\mathbf{p})
+
\frac{1}{2\beta}
\left(
\Psi_L^2 \theta_L^2
+
\Psi_R^2 \theta_R^2
\right)
\right)
\right],
\end{equation}
where $\beta = 1/T$ (we set $k_B=1$) and $H(\mathbf{x},\mathbf{p})$ is the Hamiltonian of the isolated pinned chain,
\begin{equation}
H(\mathbf{x},\mathbf{p})
=
\sum_{i=1}^{N} \frac{p_i^2}{2m}
+ V(\mathbf{x}),
\end{equation}
with $V(\mathbf{x})$ defined in Eq.~\eqref{eq:pot_pin}.
Denoting by $G(\Gamma)$ the vector field generating the dynamics in the extended phase space
$\Gamma = (\mathbf{x},\mathbf{p},\Psi_L,\Psi_R)$, the phase-space divergence reads
\begin{equation}
\nabla_\Gamma \cdot G
=
- \Psi_L - \Psi_R,
\end{equation}
since all bulk degrees of freedom evolve 
under Hamiltonian dynamics
and only the thermostatted momenta contribute to phase-space
contraction.
Using the explicit form of $f_0$ and the equations of motion, the gradient of the logarithm of the equilibrium density is
\begin{equation}
\nabla_{\Gamma} \ln f_0
=
-\beta \nabla_{\Gamma} H
-
\Psi_L \theta_L^2 \hat{\Psi}_L
-
\Psi_R \theta_R^2 \hat{\Psi}_R,
\end{equation}
where $\hat{\Psi}_{L,R}$ denote unit vectors along the thermostat directions.
After straightforward algebra, and exploiting the Hamiltonian structure of the bulk dynamics, the dissipation function
reduces to contributions arising solely from the thermostatted degrees of freedom,
\begin{equation}
\Omega^{(0)}(\Gamma)
=
\Psi_L
\left(
\frac{p_1^2}{m T_L} - \frac{p_1^2}{m T}
\right)
+
\Psi_R
\left(
\frac{p_N^2}{m T_R} - \frac{p_N^2}{m T}
\right).
\label{eq:omega_fput}
\end{equation}
When $T_L = T_R = T$, the system is unperturbed and the dissipation function vanishes identically, as required by
equilibrium~\cite{evans2016fundamentals}. This explicit expression allows the TTCF formalism to be applied directly to
the pinned anharmonic chain, expressing the nonequilibrium response of observables, such as the heat flux $J$, in terms of
equilibrium time-correlation functions computed with the density $f_0$.

\textcolor{black}{In the macroscopic description of continuous media, heat transport is characterized by the local heat flux density, which, according to Fourier’s law, is proportional to the temperature gradient.} In this framework, the thermal conductivity $\kappa$ quantifies the material’s ability to conduct heat and is defined by the ratio between the stationary heat flux and the imposed temperature gradient. In finite one-dimensional systems subject to a temperature difference $\Delta T$ across a length $N$, this relation is commonly expressed in the form
\begin{equation}
\kappa = \frac{N}{\Delta T} J,
\end{equation}
where $J$ denotes the steady-state heat flux. Establishing the scaling of $\kappa$ with the system size thus provides direct insight into the nature of heat transport—diffusive, anomalous, or ballistic—in low-dimensional systems. While several equivalent microscopic definitions exist~\cite{lepri2003thermal,giberti2010anomalies}, in this work we adopt the following expression:
\begin{equation}
J =
\frac{1}{N_B}
\sum_{i \in \mathrm{bulk}}
\dot{x}_i \,
\bigl( r_{i+1} - r_i\bigr)\;,
\end{equation}
where the sum runs over the $N_B$ bulk particles of the chain.
Equilibrium corresponds to the case in which the two thermostats are set at the same temperature,
$T_L = T_R = T$.

\section{\large Numerical Results }
In this section, we present the main experimental results for the Lorentz gas and the pinned chain.
In both systems, the equations of motion are integrated using a fourth-order Runge–Kutta (RK4) scheme. For the Lorentz gas, the integration time step is set to $10^{-4}$, while for the one-dimensional chains it is fixed to $5 \times 10^{-3}$. All the remaining \textcolor{black}{parameters} are kept constant throughout the simulations; in particular, the thermostat coupling parameters are set to $\theta_{L,R}=2$.\\
Concerning the sampling of the initial distribution $f_0$ required by the TTCF formalism, the Lorentz gas allows for a straightforward uniform sampling of initial conditions. In contrast, for the  chain we construct a buffer of equilibrium initial conditions obtained from preliminary equilibrium simulations, generating additional samples when necessary.\\
All simulations were performed on CPU architectures. The TTCF computations were parallelized using MPI and executed on the high-performance computing facilities of Politecnico di Torino (HPC@Polito), Istituto Italiano di Tecnologia (HPC@Franklin), and Politecnico di Milano (HPC@MOX). All experiments were conducted using the implementations available at
\url{https://github.com/Davidedaca/Lorentz} and
\url{https://github.com/vdiflorio/1Dchain}.

\begin{figure}[h]
    \centering
    \includegraphics[width=0.7\textwidth]{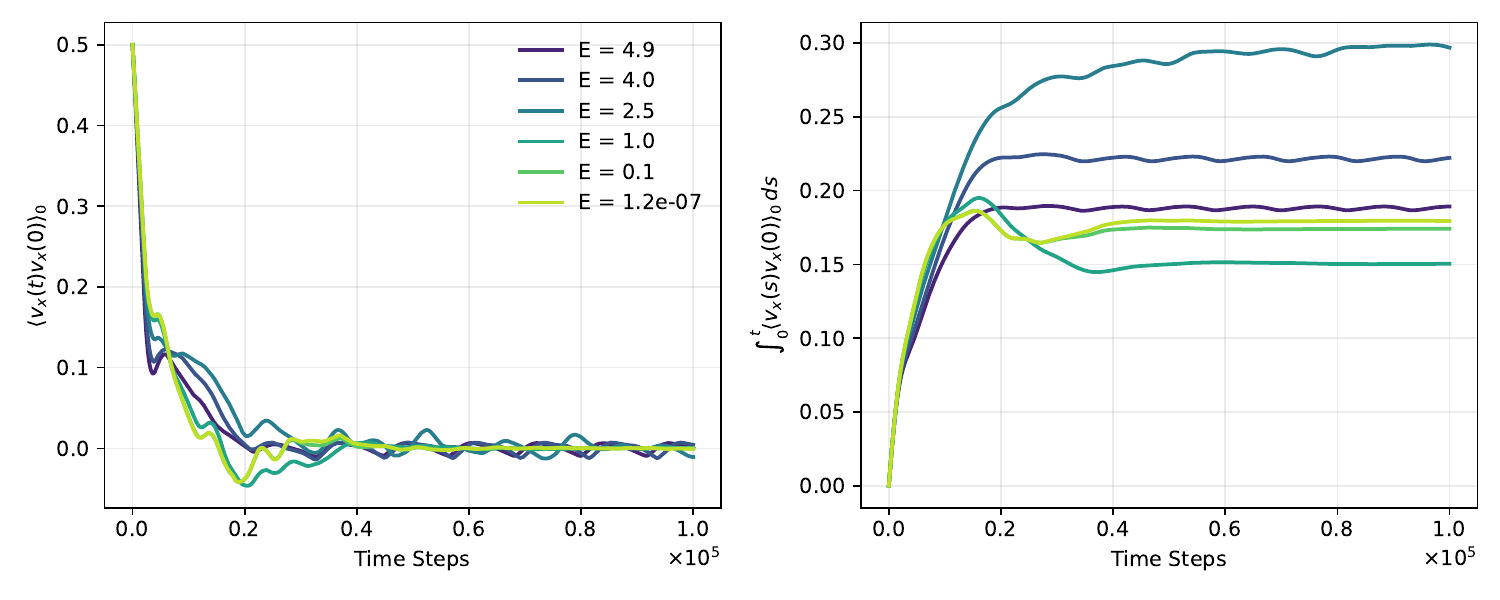}
    \caption{Transient time correlation function (TTCF). $\langle v_x(t)\,v_x(0)\rangle_0$ (left panel) and its time integral
$\int_0^t \langle v_x(s)\,v_x(0)\rangle_0\,\mathrm{d}s$ (right panel) for increasing values of the external field $E$ and for an ensemble of $10^6$ particles.
Curves are ordered \textcolor{black}{from bright to dark} colors according to increasing field strength. The cumulative integral approaches a plateau as expected from response theory. The periodic oscillations at higher fields indicates a residual correlation with initial conditions, due to finite size effect. }
    \label{fig:ttcf-Lorentz}
\end{figure}
\begin{figure}[h]
    \centering
    \includegraphics[width=0.65\textwidth]{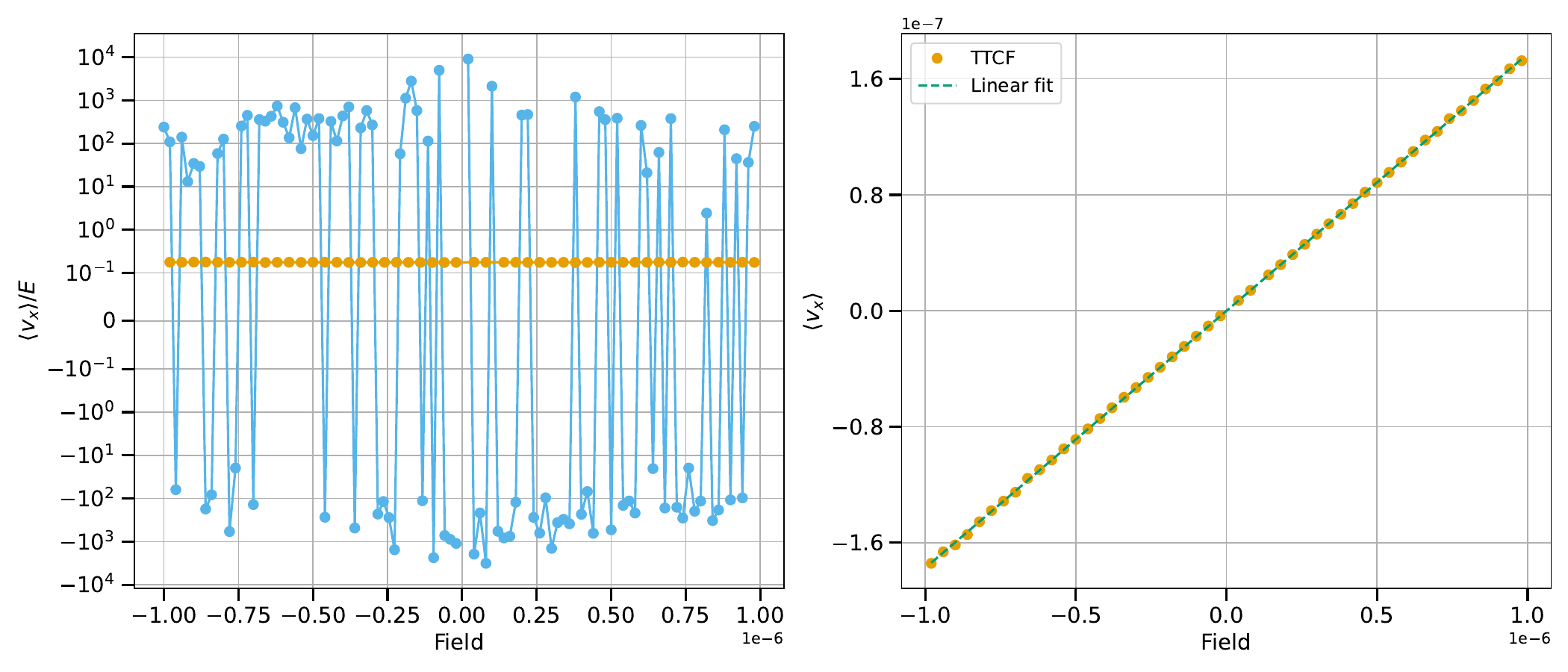}
    \includegraphics[width=0.3\textwidth]{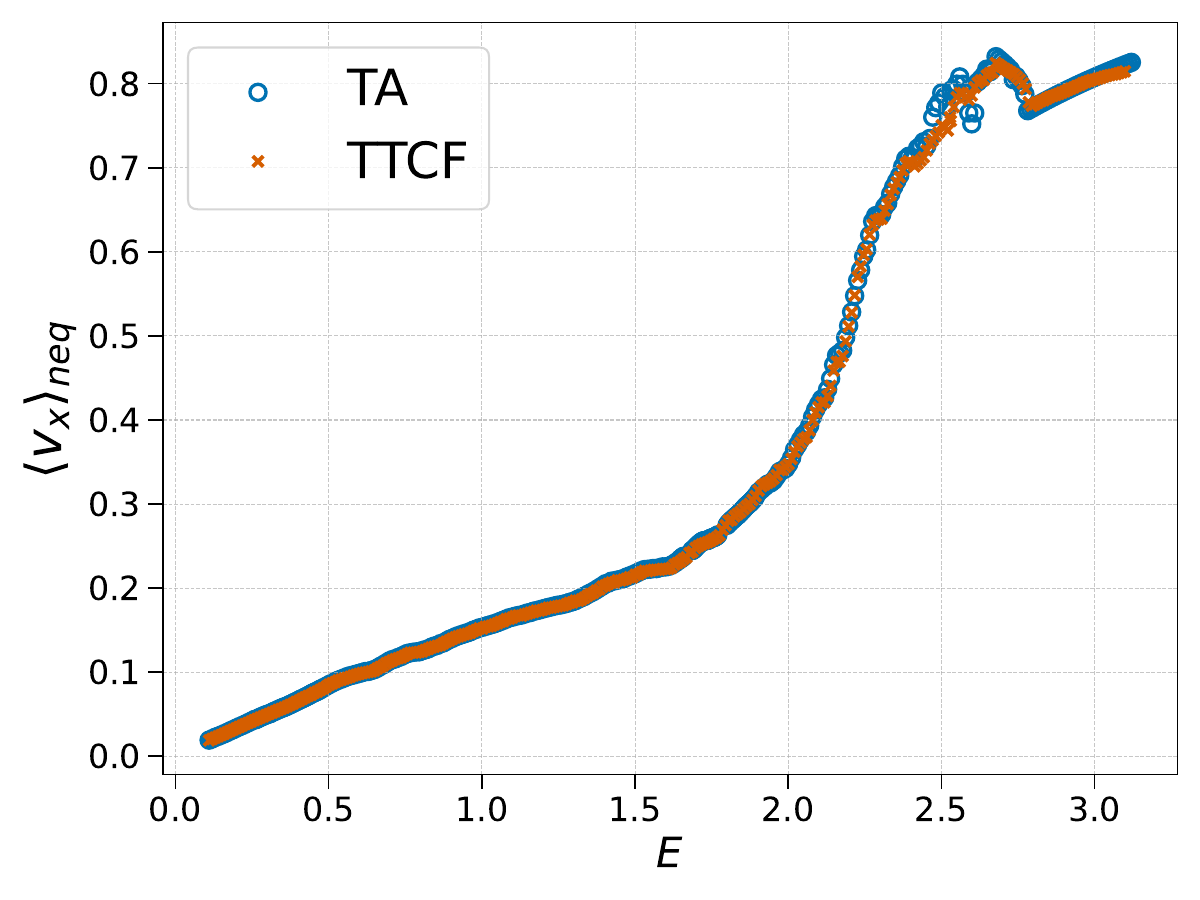}
    \caption{
{\it Left}: Comparison between time averages and TTCF predictions in the linear-response regime (small external fields).
In this regime, time averages display large fluctuations 
as the signal is small compared to the intrinsic fluctuations of the system.
{\it Center}: Zoom on the TTCF results in the linear regime, showing that, despite the small magnitude of the flux,
the response remains linear in the applied field. The value of the fitted response coefficient is $0.17$ (intercept $7\cdot10^{-11}$) and $R^2=1$ for the fit.
{\it Right:} Comparison between time averages (TA) and TTCF predictions in the nonlinear-response regime (strong external fields).
A very good agreement between TA and TTCF is observed over most of the explored field range.
Noticeable discrepancies arise in the region around $E_x \simeq 2.5$, where a 
{\color{black} splitting}
of phase space occurs,
leading to the emergence of periodic orbits with zero flux.}
    \label{fig:linear_regime}
\end{figure}
\begin{figure}[h]
    \centering
    \includegraphics[width=0.8\textwidth]{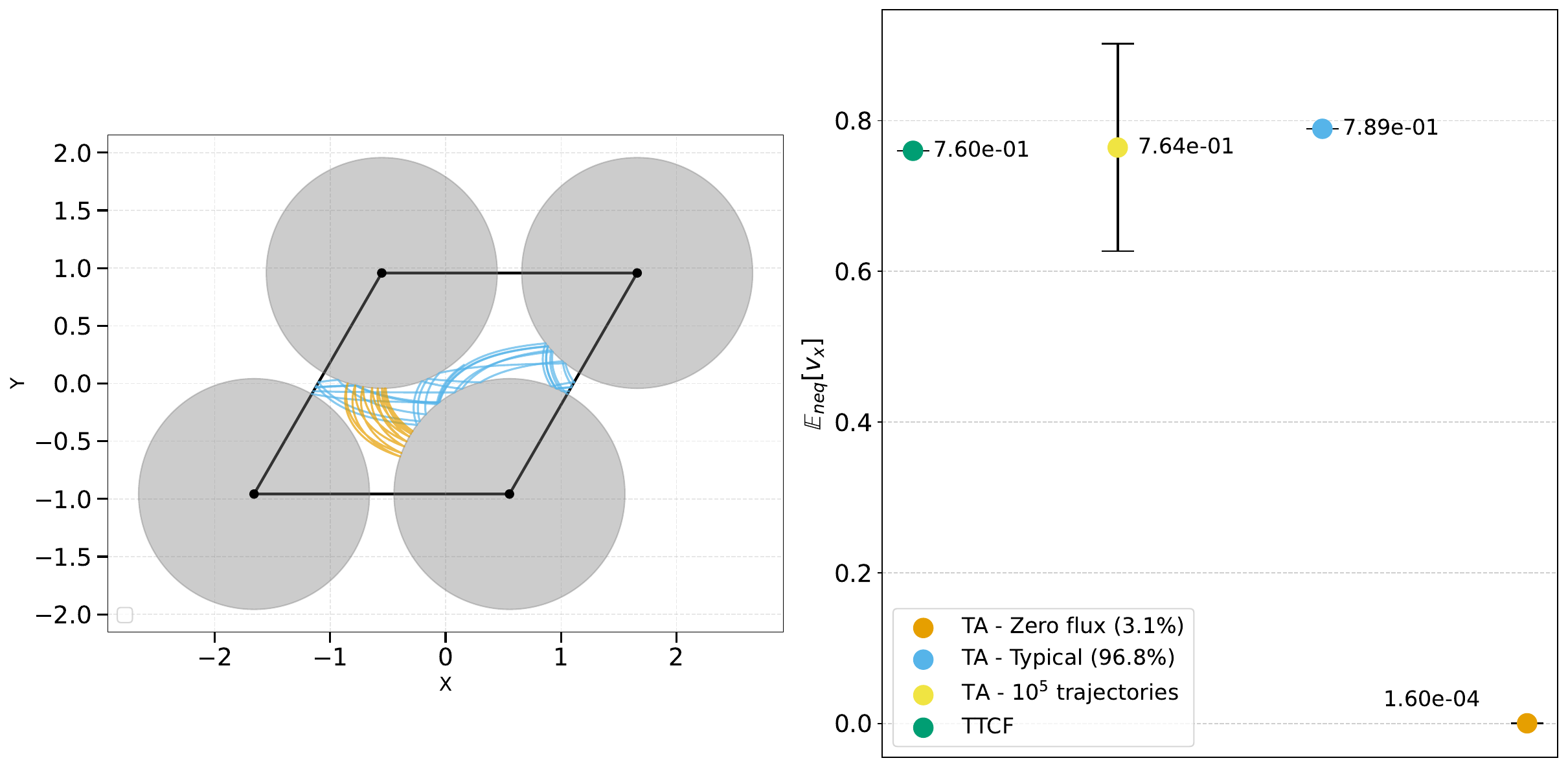}

\caption{{Nonequilibrium Lorentz gas.}
{\it Left}: Examples of typical trajectory (sky blue) and zero-flux island trajectory (\textcolor{black}{orange}). {\it Right}: Comparison of the non-equilibrium momentum flux $\mathbb{E}_{neq}[v_x]$ for a driving field $E_x = 2.528$ obtained from different methods: transient time correlation function (TTCF, \textcolor{black}{green}), time average (TA) over positive-flux trajectories (sky blue), TA over zero-flux trajectories (\textcolor{black}{orange}), and TA computed from $10^5$ trajectories (yellow). Error bars indicate the standard deviation for TA datasets and semidispersion for TTCF. Percentages in the legend denote the fraction of trajectories contributing to each TA category.
}
    \label{fig:island}
\end{figure}

\paragraph{Lorentz Gas.} Our first investigation concerns the behavior of the TTCF and of its time integral. We recall that, for the Lorentz gas, given the expression of the flux in~\eqref{eq:flux_lorentz}, the TTCF reads
\begin{equation}
\label{eq:ttcf_lorentz}
    \mathbb E_0 [\Omega^{(0)}(J\circ \Phi^s)]=\mathbb E_0 [v_x(t)E_xv_x(0)]\; .
\end{equation}
In Fig.~\ref{fig:ttcf-Lorentz} we report the TTCF and its time integral for different values of the external field $E_x$. The integrands display qualitatively similar behavior, in particular in terms of their decay properties. 
However, 
{
at the high values of $E_x$, we observe} residual periodic oscillations. These are not a contradiction with $\Omega t$-mixing, but they can be interpreted as an effect of finite size of the ensemble. In fact, for high intensity of the field the phase space region occupied by \textcolor{black}{trajectories} shrinks, 
{\color{black} undergoing various transitions, till they eventually become}
periodic, 
{\color{black} when dissipation dominates the source of chaos,}
\textcolor{black}{cf. Ref.} \cite{lloyd1994breakdown}. Once converged on the attractor, any residual correlation with initial condition, due to finite size, cannot decay perfectly \cite{giberti2007temporal}. However, given the symmetry of the initial condition, such oscillations do not drift. \\
The second investigation we present regards the linear regime. In Fig.~\ref{fig:linear_regime} on the left we plot the results of the response coefficient for $E_x\in[-10^{-6},10^{-6}]$; the expected result is a constant in the linear regime. We observe that while the coefficient computed with TTCF is very stable independently from the value of the field, the result of time average shows oscillations of order $10^3$. We stress how the comparison is fair in term of computational cost since for TTCF we employed an ensemble of $10^6$ particles for $10^5$ time steps, while for the time average we evolved a single particle for $10^{11}$ time steps. The goodness of the measure of the response coefficient is confirmed by the linear fit of $\mathbb{E}[v_x]$ computed with TTCF in the center plot. \\
This showcases the first advantage of TTCF over time average: when the nonequilibrium forcing is very small, namely in the linear regime, the exact response formula is extremely more precise due to the fact that the external field appears in the dissipation function, see \eqref{eq:ttcf_lorentz}. On the other hand, the time average computation must extract the signal from the perturbed dynamical evolution in time, since the time integrated flux does not depend on $E$ explicitly. Such advantage of TTCF over time average in the linear regime was already highlighted for other dynamical systems, see for instance \cite{MAFFIOLI2024109205}.
\\
With regards to the nonlinear regime, hence for higher values of the forcing field, we display in the right plot in Fig.~\ref{fig:linear_regime} the comparison between time average and TTCF for $E_x\in [0.1,3.1]$. The agreement is perfect except for a region around $E_x\simeq 2.5$. Besides such interval, that we will discuss in the following paragraph, we demonstrated the effectiveness of TTCF also in the nonlinear regime, compared to time average as benchmark.\\
{ Let us now} discuss the disagreement {between direct time average and TTCF results} around $E_x\simeq 2.5$. The reason of the mismatch is the structure of the phase space around that value, as previously investigated by Lloyd et al. in~\cite{lloyd1994breakdown,lloyd1995nonequilibrium}. 
In fact, around that value the phase space splits in two disconnected regions containing two different classes of trajectories (see Fig.~\ref{fig:island}).
{ For instance, let us take $E_x=2.528$, which is a point in an interval in which the same phenomenon takes place}. The first class of trajectories (in light blue) are the {\it typical} ones, {in the sense that} they occupy around $97\%$ of the phase space: they are characterized by a positive net flux. 
From a physical point of view, they are { open} trajectories that transport mass along the direction of the external field $E_x$, from left to right in the left plot. On the other hand,  about $3\%$ of the trajectories (shown in orange) are characterized by {\it zero flux} of mass, as they remain trapped in a periodic scattering motion. The authors of~\cite{lloyd1995nonequilibrium} identified these trajectory as belonging to a neutrally stable island in phase space.  \\
This peculiar behavior of the Lorentz gas provides an ideal test case for comparing TTCF and time averages in a nontrivial setting where global ergodicity is absent, as the phase space is partitioned into multiple disconnected {invariant sets}. 
To prove the critical advantage of TTCF over single-trajectory time average in this setting, we simulated $10^5$ trajectories with random initial conditions.  The resulting time-averaged fluxes are summarized in a boxplot (right panel of Fig.~\ref{fig:island}), where the average-flux is shown in yellow and the error bars represent the standard deviation over the full population.
Moreover, we separated the trajectories with zero flux (in orange) from the typical ones (in sky blue), and we computed the corresponding average flux and standard deviation for each subset. Finally, these three estimates are compared with the TTCF prediction reported in the right plot in Fig.~\ref{fig:island}. \\
\textcolor{black}{The first comment is about the standard deviation. As expected, error bars associated with TA remain very large, since the ensemble consists of two non-negligible populations characterized by markedly different flux values. Moreover, in a single realization there is approximately 3\% probability of sampling a zero-flux trajectory. From a statistical perspective, increasing the size of the ensemble would reduce the standard error of the mean, but the standard deviation remains large due to the bimodal nature of the distribution and the loss of ergodicity in this regime.
On the other hand, the numerical advantage of TTCF is evident: the measured average flux is affected by a small error, computed as the semidispersion in time on the exact response cumulative integral. Importantly, the error bars here reflect the actual uncertainty for the fully parallelizable TTCF ensemble rather than assuming a Gaussian distribution as in the standard error of the mean.} Moreover, in terms of computational cost the comparison is again in favour of TTCF: for TA we run an ensemble of $10^5$ trajectories of $10^{11}$ time steps, against $10^{6}$ trajectories for $10^{5}$ time steps in TTCF computation. \\
To summarize the two main results of the present section: first, TTCF provides substantially higher precision at fixed computational cost in the linear-response regime, where the signal-to-noise ratio is typically very small, due to the explicit dependence of the { dissipation function} on the external forcing. Second, { using the TTCF formula to compute the first two moments of the current can reveal even subtle ergodicity breakings, that may correspond to phenomena like phase transitions and hysteresis, not captured by TA.}

\begin{figure}[h!]
    \centering
    \includegraphics[width=0.9\textwidth]{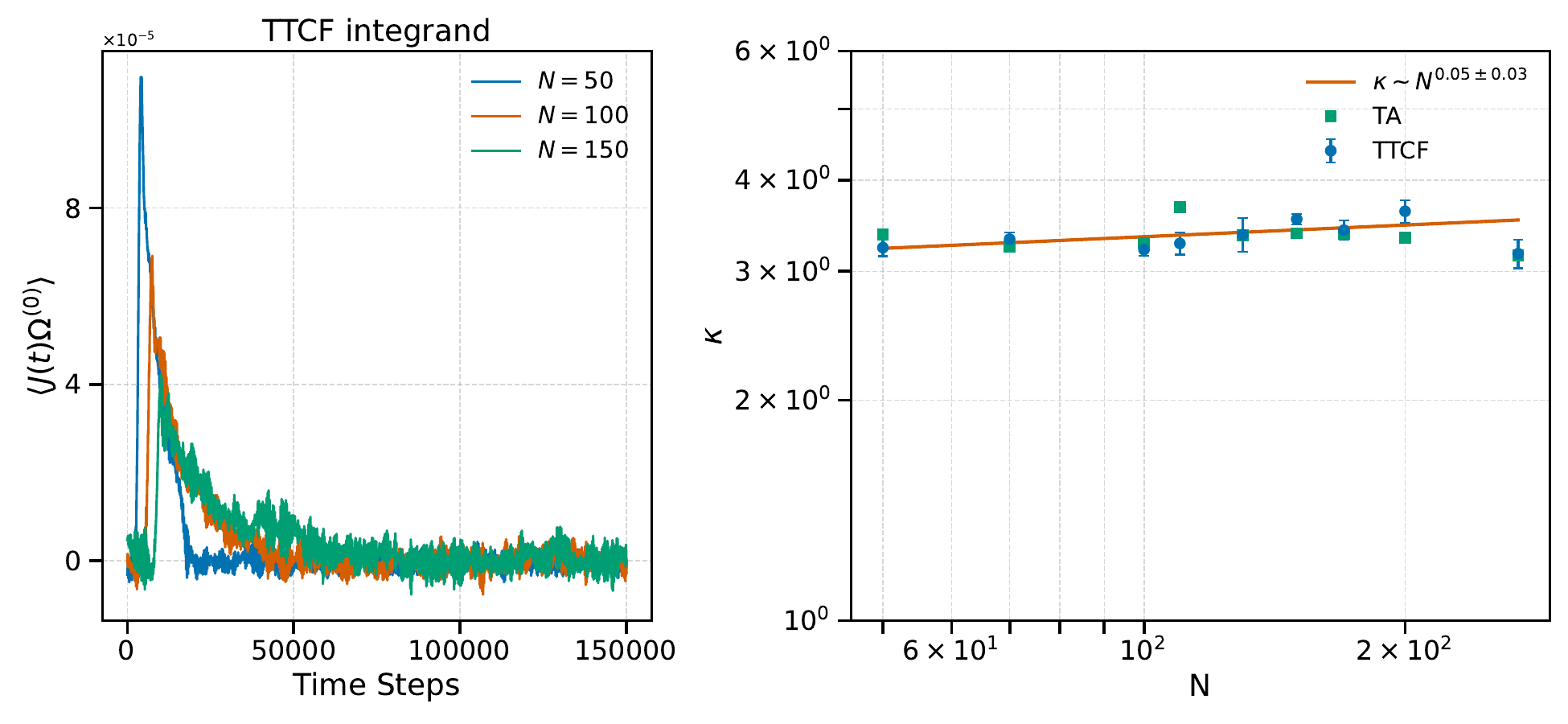}
\caption{{Pinned anharmonic chain.} {\it Left}: Transient time correlation function (TTCF)
$\langle J(t)\,\Omega^{(0)}\rangle_0$ for several chain lengths at a temperature gradient $\Delta T/N = 10^{-3}$.
{\it Right}: Comparison between TA and TTCF predictions in the linear-response regime (small temperature gradient).
In this regime, the system exhibits normal heat transport. }
    \label{fig:linear_regime_FPUT}
\end{figure}

\paragraph{Anharmonic chain} 
We next investigate the transport properties of the 
anharmonic chain as introduced in section~\ref{sec:pinned}. We stress that in this example we did not fully explore the computational scalability of TTCF, hence the comparison with TA in terms of computational cost will not be the focus of the present section.  
All simulations are performed at fixed gradient $\Delta T/N=\{0.001,1\}$ in adimensional units. In particular, the temperature of the left thermal reservoir is fixed at $T_L = 1$, while, in the nonequilibrium regime, the temperature of the right reservoir $T_R$ is adjusted according to the chain length $N$ in order to maintain a constant temperature gradient. The bulk is defined as the central 70\% of oscillators in the chain. The objective is to compute the thermal conductivity $\kappa$ as a function of $N$. In the linear-response regime, the pinned  chain exhibits normal heat conduction, i.e., a size-independent conductivity, in agreement with previous studies~\cite{lepri2003thermal,aoki2000bulk,hu2000heat}. \\
In Fig.~\ref{fig:linear_regime_FPUT} we consider the case of smaller gradient 
$\Delta T/N=0.001$. In the left panel, we plot the TTCF for different values of $N$ and for an ensemble of $10^7$ chains. A clear spike marks the onset of the nonequilibrium phase, whose position is delayed as the chain length increases. This behavior is expected: since wave propagation in the pinned chain occurs with 
finite, $N$-independent speed,  the time to propagate energy fluctuations between the thermostats is proportional 
to the chain length. Unlike simpler systems such as the Lorentz gas coupled with an isokinetic thermostat, thermal fluctuations are here much larger, even after the transient. Nevertheless, it is possible to extract the exact response after the transient, and associate to it error bars as the standard deviation of the cumulative integral considered as a stationary time series.

\begin{figure}[h!]
    \centering
    \includegraphics[width=0.9\textwidth]{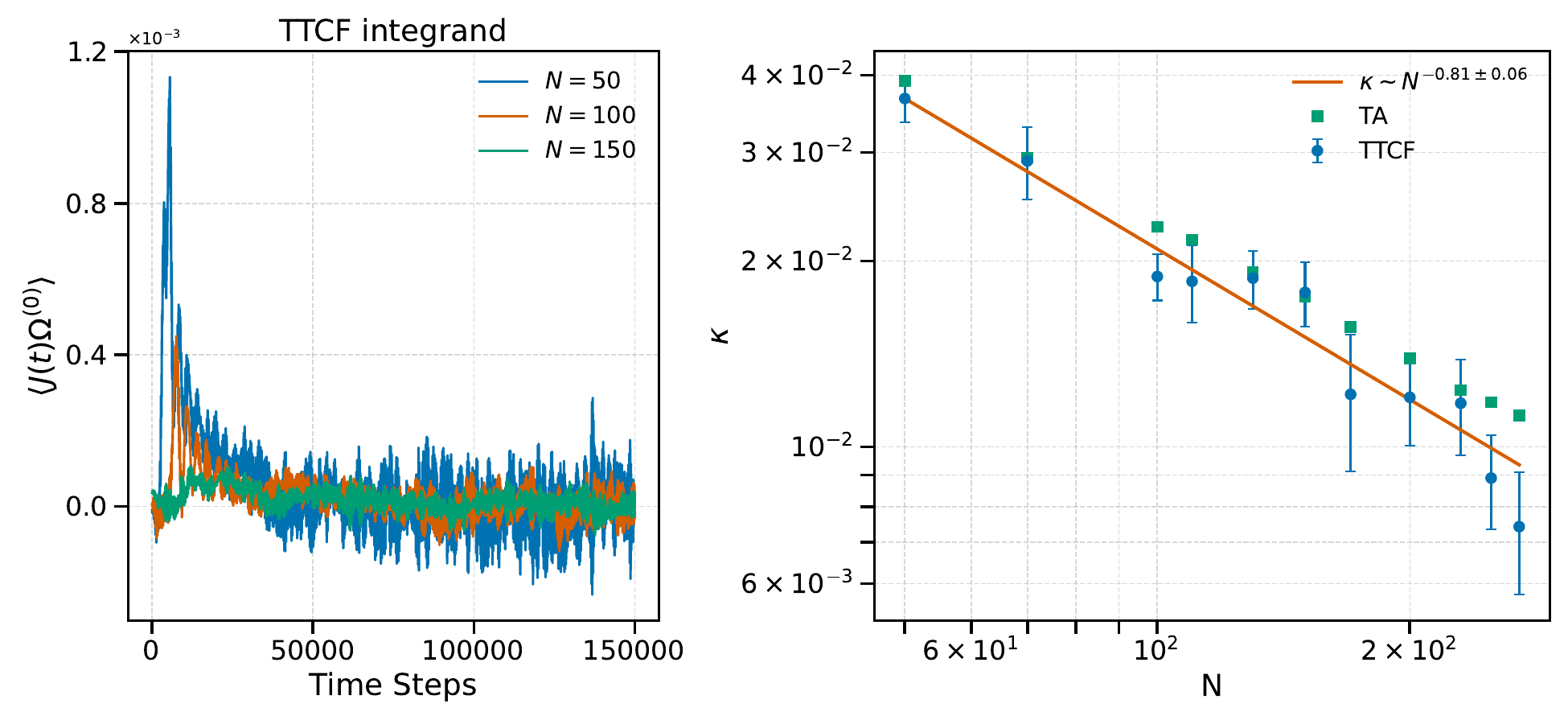}
\caption{{Pinned anharmonic chain.}
{\it Left}: Transient time correlation function (TTCF)
$\langle J(t)\,\Omega^{(0)}\rangle_0$ for several chain lengths at a temperature gradient $\Delta T/N = 1$.
{\it Right}: Comparison between TA and TTCF predictions in the nonlinear-response regime.
In this regime, deviations from linear response appear and the system displays anomalous heat transport.
}
    \label{fig:nonlinear_regime_FPUT}
\end{figure}
\begin{figure}[h!]
\centering
\begin{minipage}{0.48\textwidth}
    \centering
    \includegraphics[width=\linewidth]{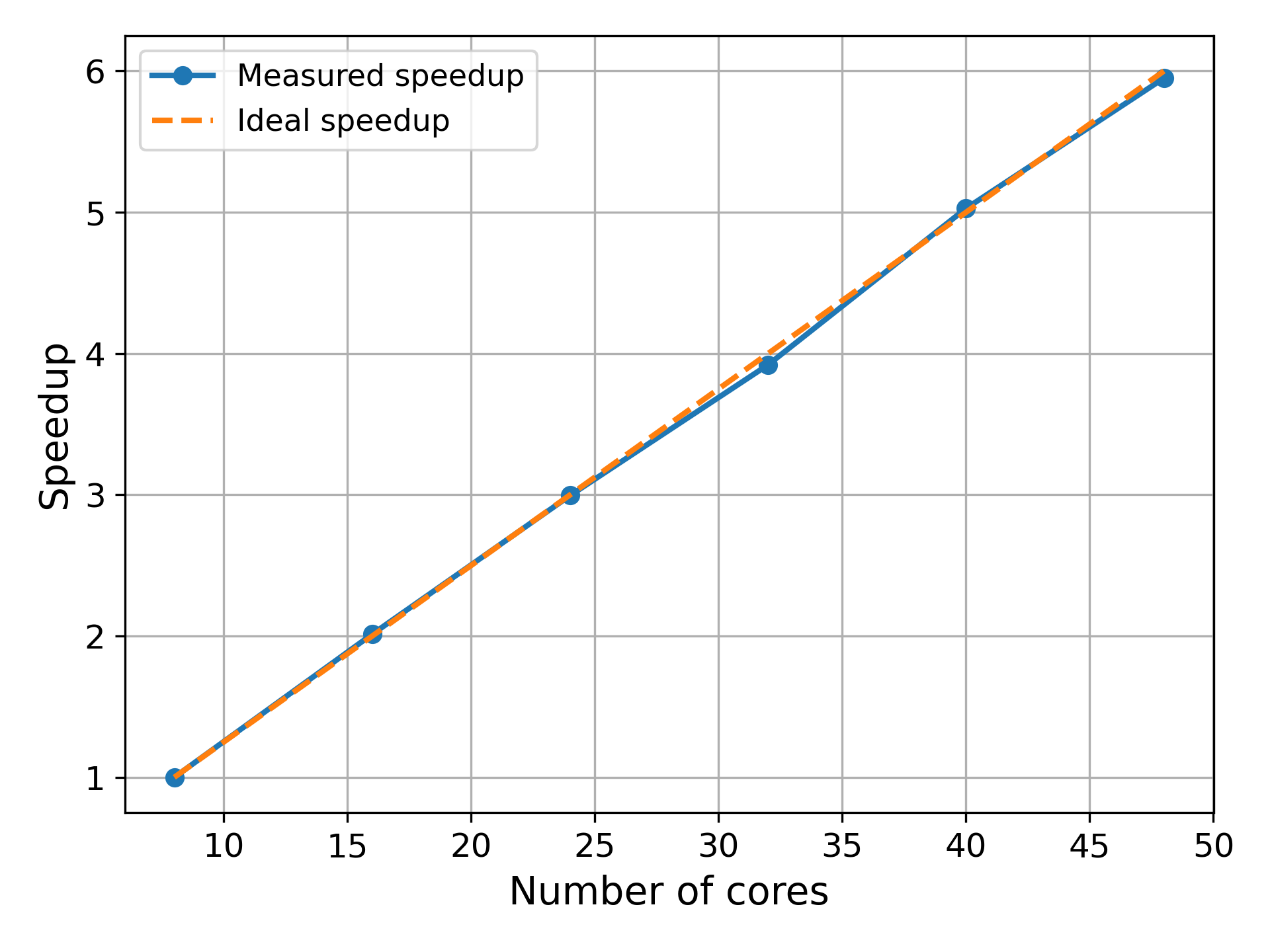}
    \caption{\textcolor{black}{Strong scaling of the TTCF approach: measured speedup (dots) compared to ideal linear scaling (dashed line).}}
    \label{fig:scaling}
\end{minipage}\hfill
\begin{minipage}{0.48\textwidth}
    \centering
    \begin{tabular}{cccc}
    \hline
    Cores & Time (s) & Speedup & Efficiency \\
    \hline
    8  & 8312.82 & 1.000 & 1.000 \\
    16 & 4129.34 & 2.013 & 1.007 \\
    24 & 2775.39 & 2.995 & 0.998 \\
    32 & 2120.45 & 3.920 & 0.980 \\
    40 & 1653.51 & 5.027 & 1.005 \\
    48 & 1397.21 & 5.950 & 0.992 \\
    \hline
    \end{tabular}
     \captionof{table}{\textcolor{black}{Strong scaling performance of the system. Speedup and parallel efficiency are computed with respect to the 8-core baseline.}}
    \label{tab:strong_scaling}
\end{minipage}
\end{figure}

In the right panel of Fig.~\ref{fig:linear_regime_FPUT}, normal heat conductivity is observed consistently in both the TTCF estimates—whose fits are compatible with a size-independent, constant thermal conductivity—and in the time-averaged simulations. For the TTCF computation, the ensemble was evolved for $5\times10^{5}$ time steps, whereas the time-averaged estimate was obtained from a single trajectory evolved for $10^{9}$ time steps. \textcolor{black}{As highlighted above, this numerical experiment is designed to demonstrate the high parallelizability of the TTCF approach, rather than to provide extremely precise quantitative results. The strong scaling behavior of the method is clearly observed in Fig.~\ref{fig:scaling}, where the wall-clock time decreases significantly with increasing number of cores. The reduction in execution time is substantially greater than the average computational time per task, indicating that the approach efficiently exploits parallel resources. The corresponding speedup and parallel efficiency are summarized in Tab.~\ref{tab:strong_scaling}, confirming near-linear scaling up to 48 cores.}

A qualitatively different behavior is observed for the case $\Delta T/N = 1$, shown in Fig.~\ref{fig:nonlinear_regime_FPUT}. For our aim, the main observation is that
the TTCF results are in full agreement with those obtained from time-averaged simulations, using the same set of \textcolor{black}{parameters} adopted in the linear regime. \textcolor{black}{More importantly, this agreement is achieved with reduced wall-clock time thanks to aforementioned parallelizable nature of TTCF, highlighting the effectiveness of the framework as a flexible and reliable tool for probing transport properties even in strongly dissipative regimes. Indeed, the purpose of the present simulations was to provide a proof of concept for the application of the TTCF framework to one-dimensional heat transport.}

Another feature of this nonlinear regime is that the conductivity $\kappa$  systematically decreases in the considered $N$ range, 
roughly as an inverse power of $N$, see right panel in Fig.~\ref{fig:nonlinear_regime_FPUT}. 
Since our aim here was to test the effectiveness of the method, we did not 
investigate this issue further. We just remark that, it is presumably due to the fact that for such large gradients 
heat transport may by strongly suppressed, a phenomenon described as negative differential thermal resistance 
\cite{li2006negative}. Indeed, such effect is 
usually explained by the fact that wave spectrum
depend on the energy density \cite{piazza2009heat} leading to a frequency mismatch and reduced transport.

\begin{figure}
    \centering
    \includegraphics[width=0.45\linewidth]{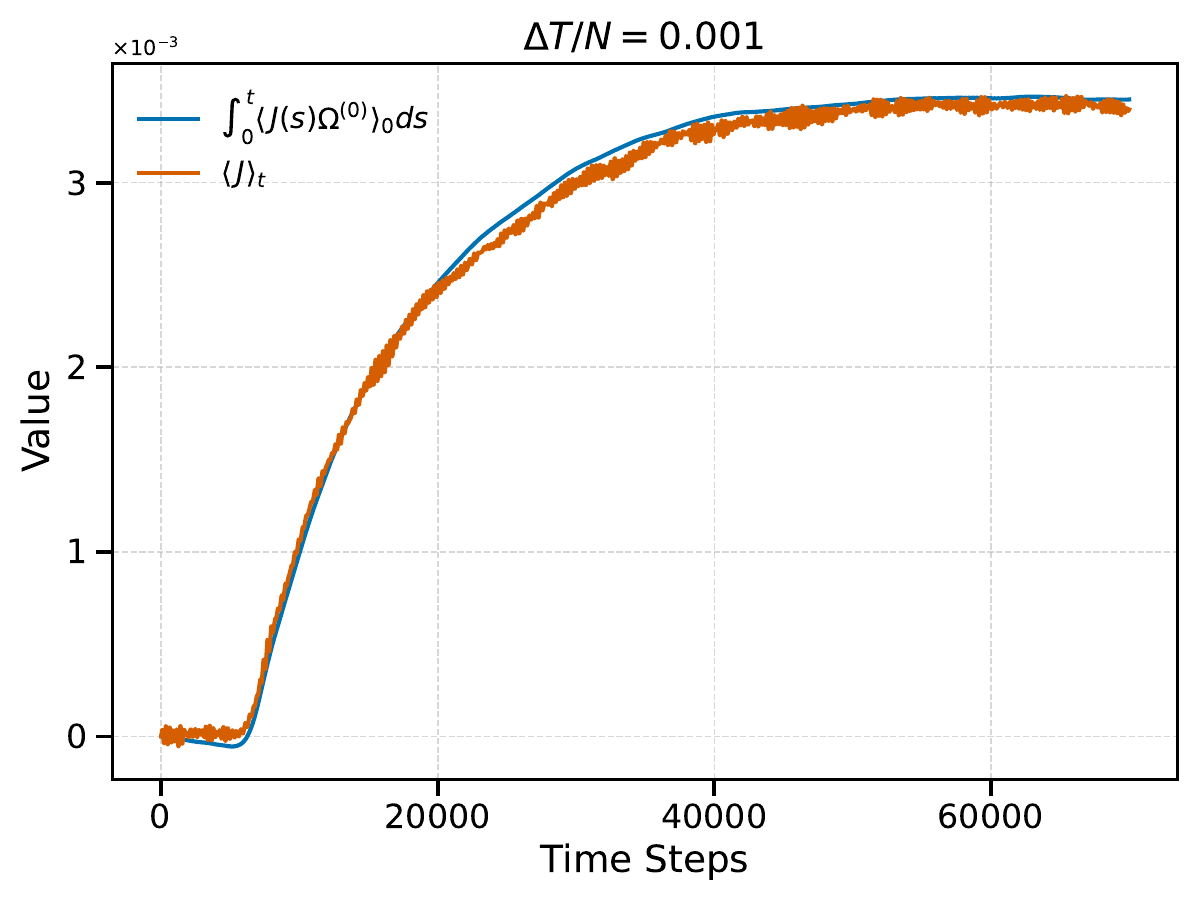}
    \includegraphics[width=0.45\linewidth]{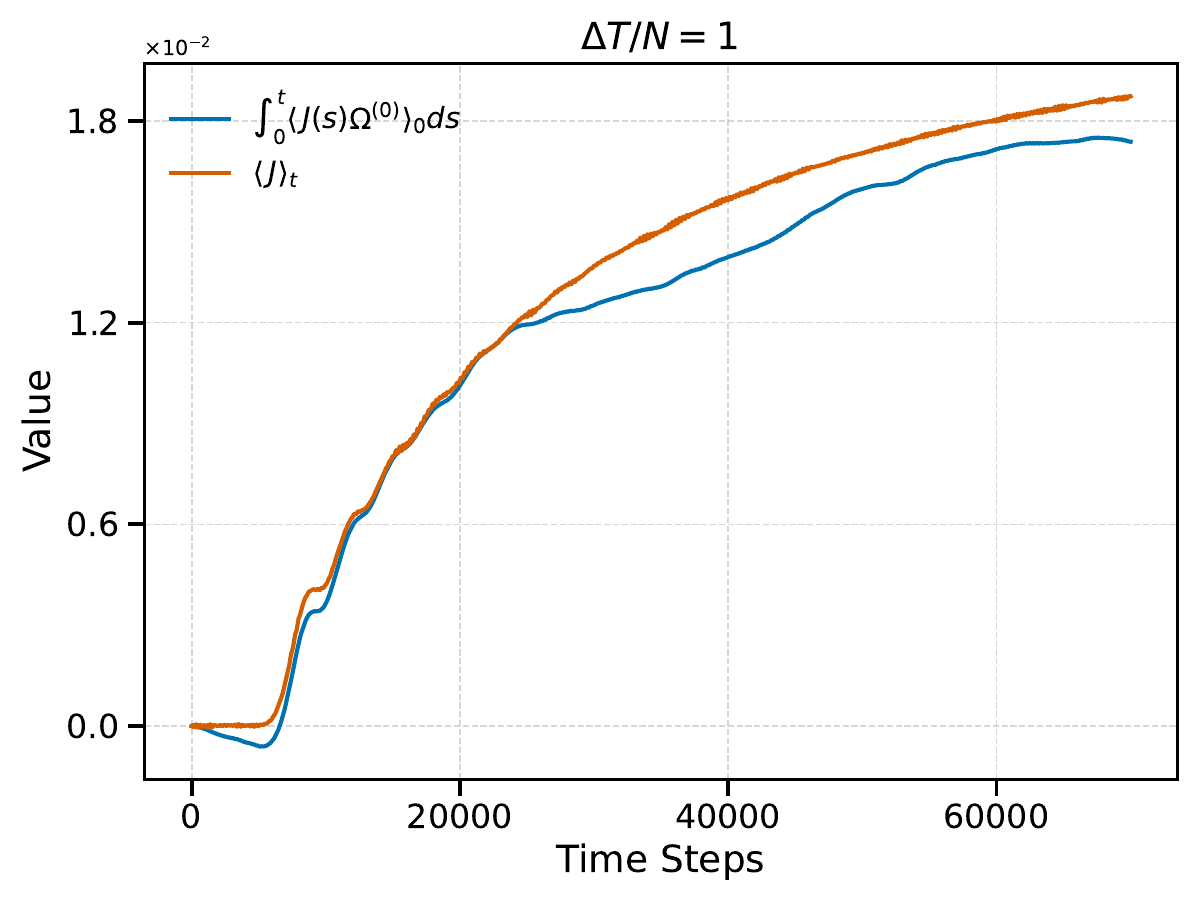}
\caption{
\textcolor{black}{Comparison between the response computed via direct ensemble averaging and via TTCF formalism for the 1D anharmonic chain at two different temperature gradients. 
{\it Left:} linear (small-gradient) regime ($\Delta T/N = 0.001$). In this case the two approaches yield very similar results within statistical uncertainty, and the advantage of TTCF is only marginal at this gradient. 
{\it Right:} larger-gradient regime ($\Delta T/N = 1$). Here the direct ensemble average provides smoother results, as the signal amplitude is sufficiently large and statistical fluctuations are less dominant.}
}
    \label{fig:ensemble_vs_TTCF_FPUT}
\end{figure}

\textcolor{black}{\paragraph{Ensemble vs. Time-Correlation Averages} To further clarify the statistical properties of our method, we compared the TTCF-based response estimate with a direct ensemble average computed from the same trajectories used for the TTCF evaluation. For the Lorentz gas, we considered up to $10^6$ trajectories for representative field strengths $10^{-3}$ and, a strongly nonlinear regime, $5$; for the 1D anharmonic chains, we used $10^7$ trajectories with corresponding temperature gradient of $10^{-3}$ and $1$. The corresponding results are reported in Figs.~\ref{fig:ensemble_vs_TTCF_Lorentz} and \ref{fig:ensemble_vs_TTCF_FPUT}.
\begin{figure}
    \centering
    \includegraphics[width=0.45\linewidth]{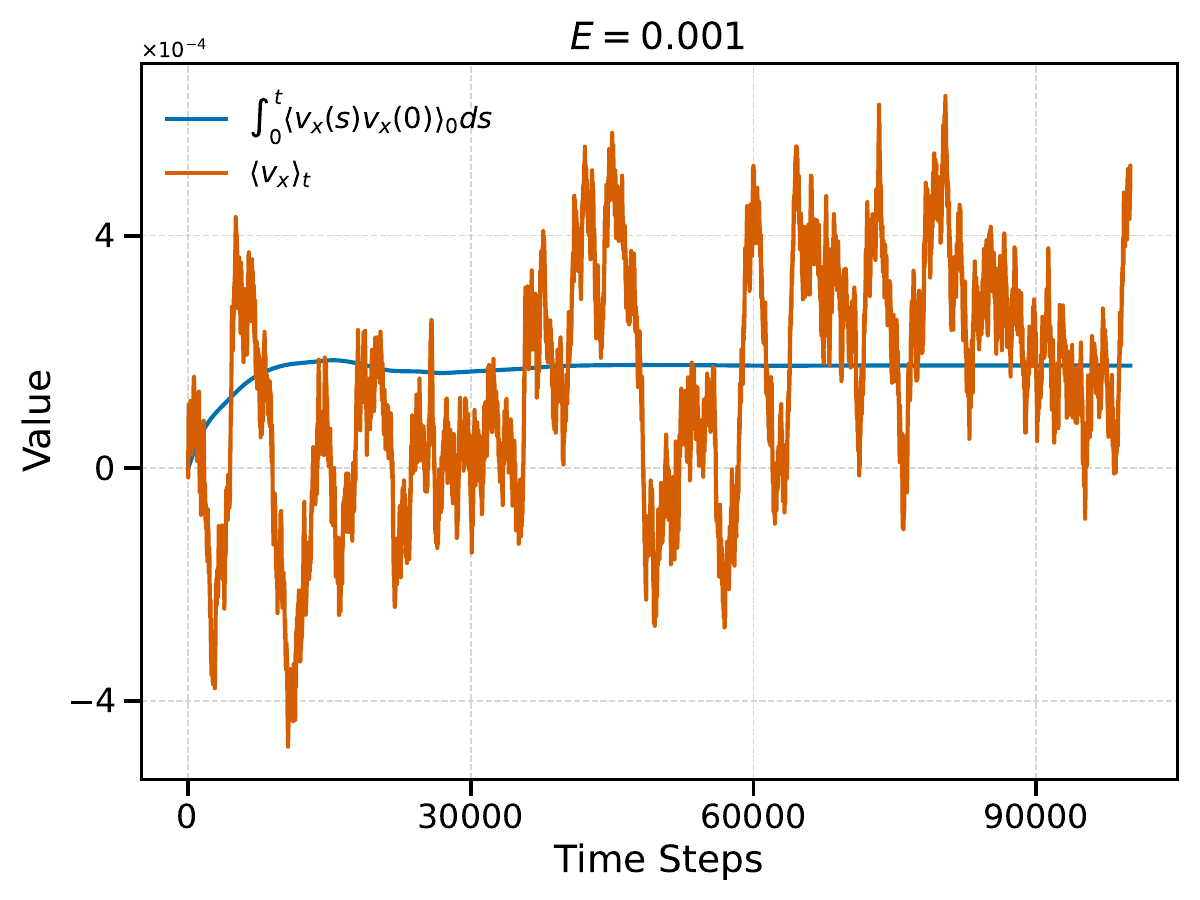}
    \includegraphics[width=0.45\linewidth]{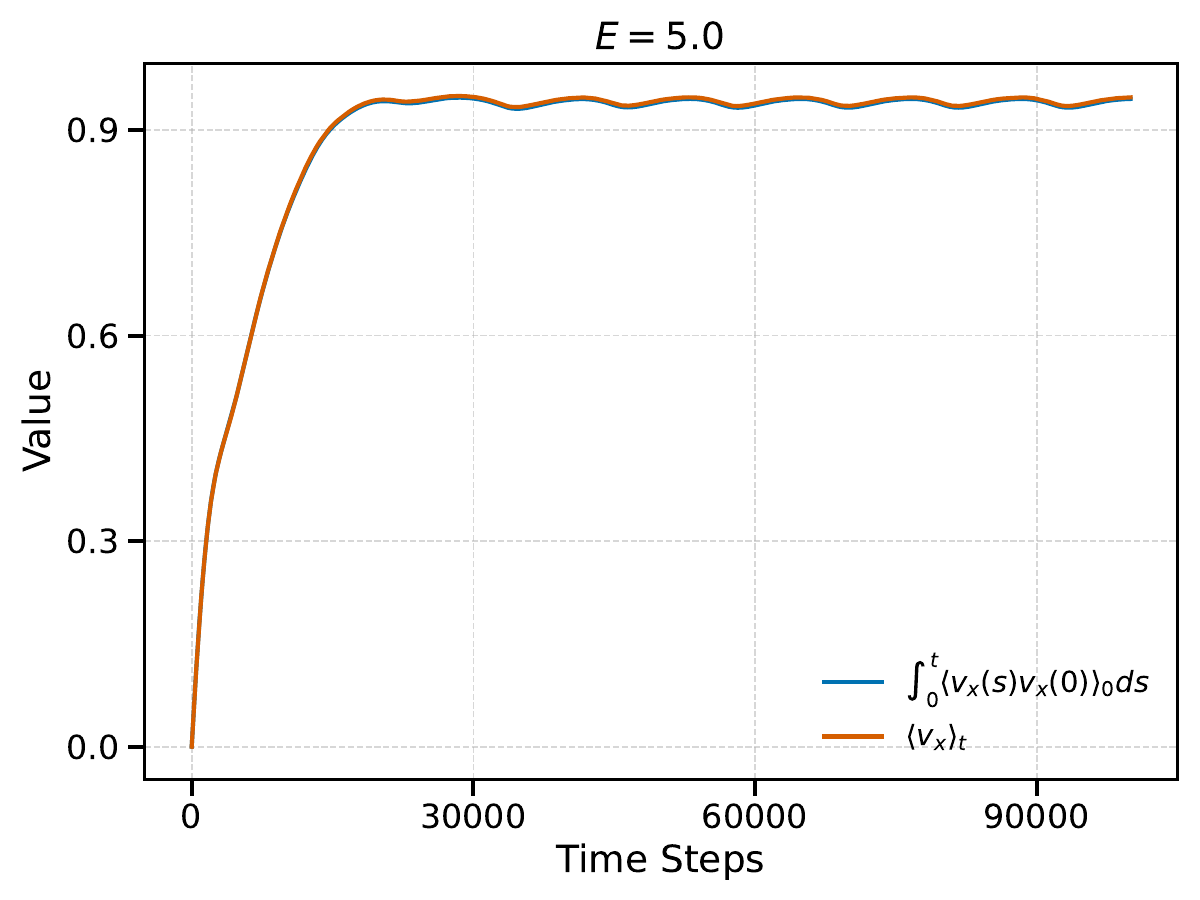}
    \caption{\textcolor{black}{
Comparison between the response computed via direct ensemble averaging and via the transient time correlation function (TTCF) formalism for the Lorentz gas at two different field strengths. 
{\it Left:} linear (small-field) regime ($F = 0.001$). In this case the TTCF approach significantly reduces statistical fluctuations, allowing for a clearer extraction of the response. 
{\it Right:} nonlinear (large-field) regime ($F = 5.0$). Here both methods produce comparable results, as the signal amplitude is large compared to statistical noise.} 
}
\label{fig:ensemble_vs_TTCF_Lorentz}
\end{figure}
Interestingly, the comparison highlights a regime-dependent behavior:
	1) {\it High dissipation (strong-field) regime}: In both the Lorentz gas and anharmonic chains, the direct ensemble average performs very well, yielding response estimates comparable, if not better, to the TTCF method. This indicates that in the strongly high dissipation regime, the signal is sufficiently strong that structural corrections from the TTCF integrand are not necessary. The good performance of the direct ensemble average also demonstrates that the signal-to-noise ratio is inherently high in this regime.
	2) {\it Linear (small-field) regime:} In contrast, the difference between the TTCF and direct ensemble averaging becomes clearly visible for sufficiently small perturbations. In this regime, TTCF reduces statistical fluctuations and enables a more precise extraction of the response, confirming that the integrand-based construction is particularly beneficial when the signal is weak.
For the Lorentz gas, this improvement is evident already at the smallest fields considered. For the 1D anharmonic chains, however, the difference is much less pronounced at the gradient $\Delta T/N = 0.001$, where the two approaches yield very close results. This suggests that, for chains, the advantage of TTCF may become more clearly visible only at even smaller gradients, where the response signal is further reduced and noise plays a more dominant role. A systematic investigation in that regime would be necessary to fully assess the gain provided by TTCF in these systems.
To the best of our knowledge, a systematic comparison of this type has not been reported previously for the Lorentz gas and for 1D anharmonic chains. The present results therefore provide a first step toward clarifying the statistical advantages of TTCF-based estimators in this class of models. Overall, the analysis helps disentangle which improvements stem from the TTCF formalism itself and which are instead attributable to ensemble averaging rather than time averaging along a single trajectory.}

\section{\large Conclusion}
In this work, we revisited the theory of the TTCF and investigated its practical performance through representative numerical examples. In particular, we compared TTCF predictions with standard time-average (TA) computations, showing that the two approaches yield qualitatively consistent results across all the considered systems and regimes.\
Our analysis of the Lorentz gas highlights several key advantages of the TTCF framework. First, TTCF proves to be computationally more efficient than time averages at fixed accuracy, requiring significantly shorter simulation times to achieve converged results. Second, TTCF displays markedly higher precision in the linear-response regime, where the signal-to-noise ratio is typically very small. In this regime, the explicit dependence of the dissipation function on the external forcing allows TTCF to suppress fluctuations and provide stable estimates even for extremely weak fields, where time averages become unreliable.\\
Moreover, TTCF remains effective in nontrivial dynamical regimes characterized by phase-space fragmentation and the absence of global ergodicity. In the Lorentz gas, we showed that TTCF correctly captures the average current even when the phase space splits into disconnected invariant sets, a situation in which single-trajectory time averages are strongly affected by large fluctuations and sampling 
biases. \textcolor{black}{Moreover, it reveals the breaking of ergodicity, allowing the analysis of the different behaviors associated with the different disjoint components of the phase space, and identifies the presence of phase transitions.}
Finally, we extended our analysis to the pinned  chain, demonstrating that TTCF also provides accurate predictions for interacting many-body systems. In this case, TTCF and time averages again show excellent agreement, confirming the robustness of the exact response formalism beyond low-dimensional chaotic models.\\
Overall, our results indicate that TTCF represents a reliable and efficient alternative to time averages for the computation of transport properties, particularly in regimes where fluctuations are large or ergodicity is weakly broken.
\textcolor{black}{These features confirms that TTCF can be a fundamental tool for the numerical study of nonequilibrium steady states in a broad class of dynamical systems.}\\
Several directions for future research naturally follow from this work. For the Lorentz gas, further investigations may address different physical observables as well as systematic variations of the scatterer geometry, in order to explore the sensitivity of TTCF to microscopic structural features. In the case of nonlinear chains, a natural extension concerns the application of TTCF to the $\alpha$- and $\beta$-FPUT models,
\textcolor{black}{which proved to be particularly challenging, and tipically require examination 
of very large sizes and/or times \cite{takatsu2024large}. In this direction, it could be interesting to investigate other possible definitions of the microscopic observable associated to the heat flux.}
Another goal is the development of variance-reduction strategies, for instance along the lines suggested in \cite{spacek2025transient}. Finally, already within the present settings, one could consider generalized TTCF formulations involving time-dependent forcing protocols or impulsive perturbations
applied at the initial time
\cite{dal2016broken,iannella2023exact}, with the aim of accelerating relaxation towards the stationary state and further improving numerical efficiency.
\subsection*{Aknowledgements}
D.C. thanks Debra Searles for very illuminating discussions. D.C. received government funding managed by the National Research Agency under the France 2030 program, reference ANR-23-IACL-0008. D.C., V.D.F. and L.R. worked under the auspices of Italian National Group of Mathematical Physics (GNFM) of INdAM. LR gratefully acknowledges funding under the project NODES, of the MUR - M4C2 1.5 of PNRR, European Union - NextGenerationEU (Grant agreement no.
ECS00000036). 
\printbibliography

\end{document}